\documentclass[%
 aip,
 amsmath,amssymb,
 reprint,%
]{revtex4-1}

\usepackage{graphicx}
\usepackage{dcolumn}
\usepackage{bm}
\usepackage{upgreek}
\usepackage{dcolumn}

\usepackage[utf8]{inputenc}
\usepackage[T1]{fontenc}
\usepackage{mathptmx}
\usepackage{etoolbox}

\makeatletter
\def\@email#1#2{%
 \endgroup
 \patchcmd{\titleblock@produce}
  {\frontmatter@RRAPformat}
  {\frontmatter@RRAPformat{\produce@RRAP{*#1\href{mailto:#2}{#2}}}\frontmatter@RRAPformat}
  {}{}
}%
\makeatother
\begin{document}

\preprint{AIP/123-QED}
\title[Improvements of readout signal integrity in mid-infrared superconducting nanowire single photon detectors]{Improvements of readout signal integrity in mid-infrared superconducting nanowire single photon detectors}

\author{Sahil R. Patel}
\affiliation{Jet Propulsion Laboratory, California Institute of Technology, 4800 Oak Grove Dr., Pasadena, CA, USA}
\email{srpatel@caltech.edu}

\author{Marco Colangelo}
\affiliation{Research Laboratory of Electronics, Massachusetts Institute of Technology, Cambridge, Massachusetts 02139, USA}
 \affiliation{Electrical and Computer Engineering, Northeastern University, Boston, Massachusetts 02115, USA.}
\author{Andrew D. Beyer}
\affiliation{Jet Propulsion Laboratory, California Institute of Technology, 4800 Oak Grove Dr., Pasadena, CA, USA}
\author{Gregor G. Taylor}
\affiliation{Jet Propulsion Laboratory, California Institute of Technology, 4800 Oak Grove Dr., Pasadena, CA, USA}
\author{Jason P. Allmaras}
\affiliation{Jet Propulsion Laboratory, California Institute of Technology, 4800 Oak Grove Dr., Pasadena, CA, USA}
\author{Emma E. Wollman}
\affiliation{Jet Propulsion Laboratory, California Institute of Technology, 4800 Oak Grove Dr., Pasadena, CA, USA}
\author{Matthew D. Shaw}
\affiliation{Jet Propulsion Laboratory, California Institute of Technology, 4800 Oak Grove Dr., Pasadena, CA, USA}
\author{Karl K. Berggren}
\affiliation{Research Laboratory of Electronics, Massachusetts Institute of Technology, Cambridge, Massachusetts 02139, USA}

\author{Boris Korzh}
\affiliation{Jet Propulsion Laboratory, California Institute of Technology, 4800 Oak Grove Dr., Pasadena, CA, USA}

\begin{abstract}
Superconducting nanowire single-photon detectors (SNSPDs) with high timing resolution and low background counts in the mid infrared (MIR) have the potential to open up numerous opportunities in fields such as exoplanet searches, direct dark matter detection, physical chemistry, and remote sensing. One challenge in pushing SNSPD sensitivity to the MIR is a decrease in the signal-to-noise ratio (SNR) of the readout signal as the critical currents become increasingly smaller. We overcome this trade-off with a new device architecture that employs impedance matching tapers and superconducting nanowire avalanche photodetectors to demonstrate increased SNR while maintaining saturated internal detection efficiency at 7.4~\textmu m and getting close to saturation at 10.6~\textmu m. This work provides a novel platform for pushing SNSPD sensitivity to longer wavelengths while improving the scalability of the readout electronics. 

\end{abstract}

\maketitle


\begin{quotation}
The mid-infrared (MIR) range is an area of rich scientific interest with several applications, from physical chemistry to astronomy~\cite{Lau2023}. In this spectral range, many strong optical absorption lines are present, each carrying detailed information about the vibrational modes of molecules, allowing their identification~\cite{Cohen2015}. The MIR range is also a growing frontier for applications in quantum sensing~\cite{Russo2022}, imaging and ranging ~\cite{Taylor2019}, and communications~\cite{Temporao2008}. As the number of applications for MIR wavelengths increases, so does the necessity to develop single-photon-sensitive detectors in this range. 
\end{quotation}

\section{\label{sec:intro}Introduction\\}  

Currently, the most common MIR detectors are based on semiconductor technology, including HgCdTe, known as MCTs~\cite{Chen2021-MCT}, and blocked impurity band (BIB) detectors~\cite{Deng2022}. However, their lack of single-photon sensitivity, excessive noise, and slow-timing performance have motivated the development of superconducting nanowire single-photon detectors (SNSPDs) for MIR applications. SNSPDs possess much faster reset times~\cite{Craiciu23}, low dark count rates~\cite{Chiles2022}, high timing resolution~\cite{Korzh2020,Allmaras2019}, along with the ability to scale towards large arrays~\cite{Oripov2023}. These attributes are especially useful when looking at applications in infrared astronomy, where the signal is very faint and requires low noise, high stability, and single photon sensitivity ~\cite{Wollman2021}, or direct dark matter detection explorations, where lower energy thresholds can unlock larger parameter space~\cite{Golwala2022}.

Previous demonstrations of MIR detection with SNSPDs have been realized using narrow wires of WSi~\cite{Verma2020, Colangelo2022}, polymorphs of NbN~
\cite{Pan2022}, NbTiN~\cite{Chang2022}, and MoSi~\cite{Chen2021}. Among these materials, WSi devices have seen the farthest incursion into the MIR by achieving unity internal detection efficiency (IDE) up to 29~\textmu m ~\cite{Taylor2023}. 
Achieving unity IDE is typically the first step towards increasing the total system detection efficiency (SDE) that is the product of the optical coupling to the detector, the absorption of photons incident on the detector, and the IDE where every photon absorbed by the wire leads to a corresponding output pulse. 

To achieve unity IDE at longer wavelengths, the typical approaches are to either reduce the cross-sectional area, use a material with a smaller superconducting gap (lower critical temperature $T_C$), or increase film resistivity (fewer quasi particles that need to be excited to generate a hot spot)~\cite{Colangelo2022}. These parameters are hard to tune independently, and a combination of these effects is typically required. Both Ref.~\cite{Colangelo2022} and Ref.~\cite{Verma2020} fabricated nanowires in the range of 40-80~nm from thin films possessing a low $T_C$ relative to previous SNSPDs optimized for the near infrared. While this technique of thin and narrow wires with low $T_C$ is effective in increasing detector sensitivity, it correspondingly lowers the switching current (the current at which the nanowire transitions from superconducting to normal state) usually to less than 1\textmu A. This generally decreases the signal-to-noise ratio (SNR), complicating the readout of voltage pulses that are proportional to the bias current supplied to the detector. As SNSPDs continue to push farther into the MIR, with narrower and lower $T_C$, novel device architectures or methods to increase the SNR will be needed. This is especially important to avoid electronic errors and enable comparable SNSPD performances in this band as in the near infrared.

In this Letter, we outline a novel device architecture involving the combination of impedance-matching tapers and superconducting nanowire avalanche photodetectors (SNAPs). With this architecture, we are able to demonstrate unity IDE at 7.4~\textmu m, in conjunction with greatly improved SNR compared to our previous results~\cite{Colangelo2022}. Individually, impedance-matching tapers have been shown to improve SNR by increasing the load of the device while still efficiently coupling to a 50~$\Omega$ readout transmission line. Moreover, they have demonstrated both reduced timing jitter and increased voltage amplitude, while preserving the fast rising (falling) edge of the pulse ~\cite{Zhu2019}. SNAPs consist of superconducting nanowires placed in a parallel configuration. They rely on the generation of a hotspot in one of the wires which then shunts bias current into the parallel wire(s) exceeding the switching current and initiating the transition into the normal state. As an effect, a proportionatly higher current is shunted to the readout electronics, resulting in a larger pulse amplitude, faster slew rate, and therefore higher SNR~\cite{Ejrnaes2007,Marsili2011}. 
 
The combined effect of impedance matching tapers and SNAPs is still relatively unexplored for the development of MIR SNSPDs as these two elements are usually designed for applications using telecom wavelengths, or shorter. Traditionally, SNAPs are coupled with a series inductor to force current into the parallel nanowires and ensure stable switching ~\cite{Ejrnaes2007}. Furthermore, for the avalanche or cascade of switching events to occur, the detector must be biased above what is known as the avalanche current. If the detector is biased below this avalanche current, the first superconducting nanowire may return to the superconducting state without shunting all of its current to adjacent sections thereby preventing the cascade of switching events from occurring ~\cite{Marsili2011}. With lower photon energies and bias currents for MIR operation, it is still unclear whether this avalanche mechanism will hold. To carry out our demonstration, we use the DC characteristic of the impedance-matching taper to replace the additional series inductance required for optimal SNAP operation. This effectively increases the load while still acting as a high-pass element, ultimately preserving the falling (rising) edge of the pulse, and improving SNR ~\cite{Zhu2019}. 

To illustrate the impact of combining SNAP and impedance matching tapers on MIR devices, we fabricated 80 nm wide WSi SNSPDs, 2-SNAP, and 3-SNAP devices. With this design, we were able to demonstrate an increase in voltage intensity proportional to the number of nanowires in the SNAP along with an increase of SNR compared to the pulses reported in Ref.~\cite{Colangelo2022}. We tested this architecture at 7.4\textmu m to ensure that adding these elements does not reduce the IDE. We also tested these devices at 10.6~\textmu m to demonstrate the current IDE saturation limit.  We believe this device architecture can be used as a template for future MIR SNSPDs, especially as SNSPDs push further into the MIR and switching currents continue to decrease.  

\section{\label{sec:level1}Device Fabrication and Design\\}

Our devices are fabricated according to a microstrip architecture, necessary to ensure good microwave propagation properties of the impedance-matching tapers. These are designed following the Hecken formalism~\cite{hecken1973optimum,mccaughan2021phidl} with an $80\,\mathrm{MHz}$ bandwidth. The fabrication started with the deposition of a 50~nm-thick Al ground (GND) plane layer using electron beam evaporation followed by a 50~nm-thick SiO$_{2}$ dielectric layer deposited with sputtering. An approximately 3~nm-thick WSi superconducting film, with a $T_c$ of 2.5~K, was then sputtered on top of the SiO$_{2}$ from a W$_{50}$Si$_{50}$ target at 130~W. Next, contact pads and top grounding layer were patterned and deposited. Finally, the nanowire devices and impedance matching tapers were patterned using electron-beam lithography with hydrogen silsequioxane (HSQ), etched in CF$_4$ plasma and capped with a thin layer of sputtered amorphous Si. An optical micrograph of the fabricated design is shown in Fig. 1.  Our nanowires have widths ranging from 40-nm to 80-nm with 2- and 3-SNAP variants for each wire width. The length of the nanowires is $20\,\mathrm{\upmu m}$. For this study we focused on the 80-nm-wide devices, fabricated with the same process on two separate dies: Die 1 and Die 2. Table \ref{table:1} summarizes the devices characterized and discussed in this paper. We chose 80-nm as they had the best yield, best scaling in switching currents (linear) between single wire up to 3-SNAP, and longest PCR plateaus at 7.4~\textmu m making them the best candidates for this study.

\begin{table}[h!]
\centering
\begin{tabular}{ |c|c|c|c| }
 \hline
  Device & Design & Switching Current & Die $\#$ \\
  \hline\hline
  A & 80nm Wire & 0.95 $\upmu$A & 1 \\
  B & 80nm 2-SNAP & 1.48 $\upmu$A & 1\\
  C & 80nm 2-SNAP & 1.52 $\upmu$A & 2  \\
  D & 80nm 3-SNAP & 2.63 $\upmu$A & 2\\
  \hline
\end{tabular}
\caption{Characterized devices and corresponding switching currents.}
\label{table:1}
\end{table}

\begin{figure}[ht]
    \centering
    \includegraphics[width=\linewidth]{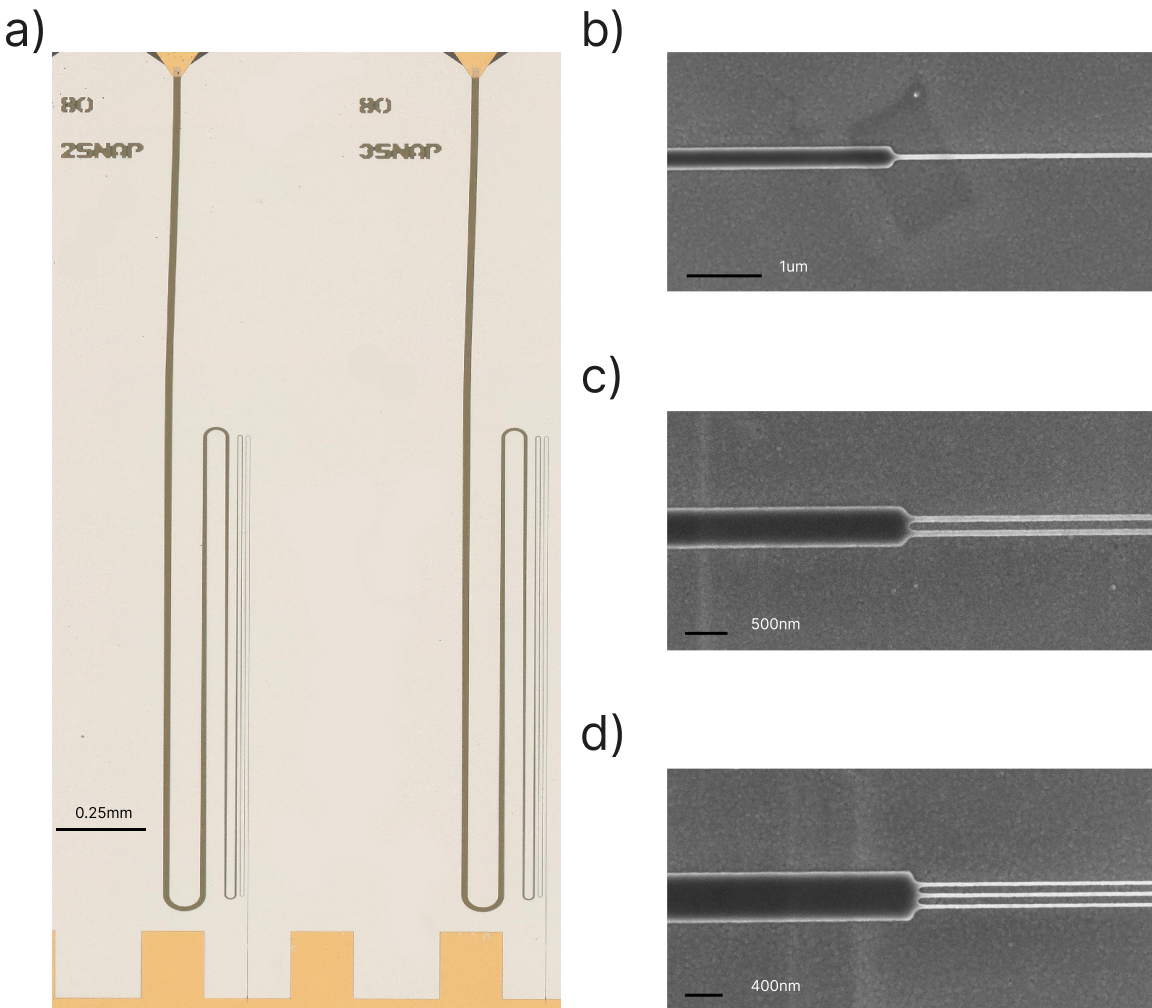}
    \caption{Micrographs of the fabricated devices. (a) Optical micrograph of completed die with impedance matching tapers and SNAPs (b-d) Scanning electron micrographs of sections of representative wire, 2-SNAP, and 3-SNAP with HSQ softmask on top. }
    \label{fig:devices}
\end{figure}

\section{\label{sec:measurementsetup}Measurement Apparatus and Setup\\} 

The devices were measured in a $^{4}$He sorption cryostat with a base temperature of 0.86~K. The devices were flood-illuminated with 7.4~\textmu m light using a BaF$_{2}$ shortpass filter and a 7.4~\textmu m narrow bandpass filter (Spectrogon) coupled to a blackbody thermal source. Additionally Device A and B were tested under 10.6~\textmu m light using the same BaF$_{2}$ shortpass filter and a 10.6~\textmu m bandpass filter (Edmunds).  Characteristic pulses were amplified at the 40~K stage using a cryogenic amplifier (Cosmic Microwave Technology CMT-LF1) along with a 120 MHz low pass filter. Using this setup we were able to measure switching currents in the hundreds of nanoamps. 

To ensure proper photon counting, pulses were first read out using an oscilloscope to determine a proper threshold for each bias point. This was done to ensure the threshold point was free from any ripples due to electronic noise that could lead to an improper measurement. Pulses were then sent to a counter for collecting data. The count rate was tuned to the single photon regime by changing the resistance of the thermal source, thereby shifting the peak of blackbody radiation. After an appropriate count rate was found, photon count rate (PCR) measurements were taken by pulsing the thermal source at 1~Hz with a 35\% duty cycle, collecting photon counts when the thermal source was on, and subtracting the dark count rate (DCR) when the source was off. To further ensure an accurate measurement, this procedure was performed for a range of threshold values for each bias current point. The final PCR curve was developed by choosing the best threshold point for each bias current point. In this way we come as close as possible to eliminating any double counting due to noise along with delayed onset currents due to thresholds being set too high.

\section{\label{sec:level2}Results\\} 
The devices selected for this study allow us to characterize the effect of coupling the SNAP architecture with impedance-matching tapers. Device A is a single wire connected to an impedance-matching taper. Device B and C have one additional parallel wire in the 2-SNAP configuration. Device D has a total of three parallel wires in the 3-SNAP configuration.  

\begin{figure} [ht]
    \includegraphics[width=\linewidth]{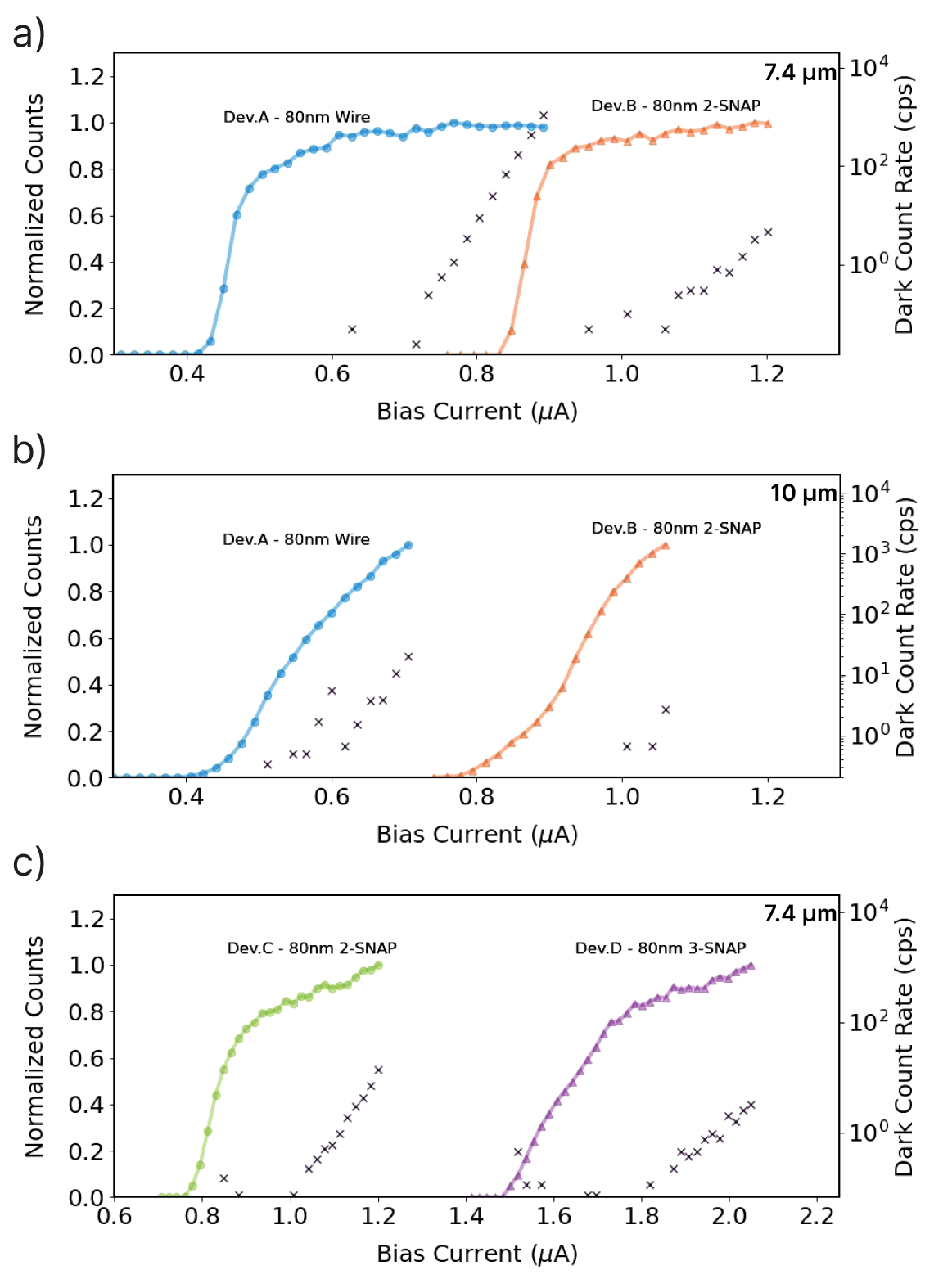}
    \caption{PCR curves taken of all four devices with dark counts shown with x -markers.(a) Device A and B PCR curves demonstrating a comparison between just impedance matching tapers in Device A and the addition of a parallel wire (2-SNAP) in Device B at 7.4\textmu m . (b) PCR taken at 10.6~\textmu m for Device A and B where both devices inflect but do not saturate. (c) Comparison of PCR curves between Device C and D demonstrating that detectors maintain sensitivity even up to a 3-SNAP configuration at 7.4\textmu m.}
    \label{fig:PCRCurves}
\end{figure}

\begin{figure}[ht] 
    \includegraphics[width=\linewidth]{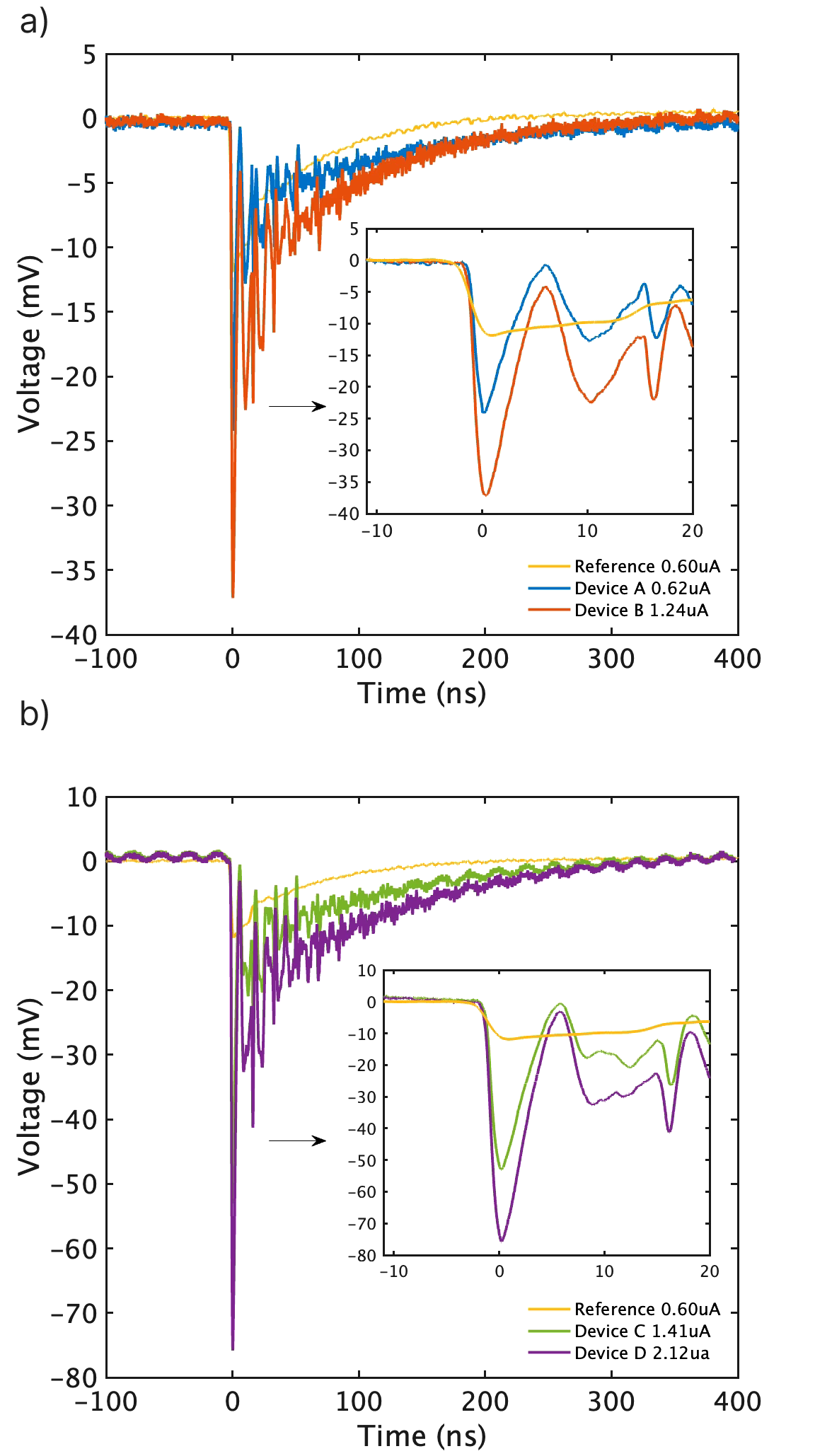}
    \caption{Pulse traces from detectors in comparison to M60 reference device. (a) shows the  80 nm bridge and 2-SNAP device from Die 1 where there is a clear improvement in slew rate and amplitude in both cases and a proportional doubling of amplitude in the 2-SNAP device. (b) shows the pulse traces from the 2-SNAP and 3-SNAP device from Die 2 in comparison to the reference device. Both show great improvement when compared to the reference device along with nice scaling in voltage amplitude.}
    \label{fig:PulseTraces}
\end{figure}

The normalized PCR curves, with the DCR are shown in Fig.~\ref{fig:PCRCurves}. All devices tested have a clear turnover in the PCR curve at 7.4~\textmu m, indicating that the addition of SNAPs and tapers does not hamper the sensitivity of the MIR detector. From the PCR curves at 7.4~\textmu m, the rise in counts begins at roughly double the bias current applied to the detector in the 2-SNAP configuration and roughly triple for the 3-SNAP configuration, as compared to the single wire. The slight uptick plateau region of the 7.4~\textmu m PCR curves at higher bias currents is attributed to the presence of longer wavelength photons and we believe could be flattened with better radiation shielding \cite{Taylor2023}. Another point of interest is the presence of a secondary plateau in the 3-SNAP near 1.6\textmu A, a typical feature found in this device acting as a pseudo 2-SNAP before transitioning to the 3-SNAP state \cite{Marsili2011}. We also tested Device A and B at 10.6~\textmu m to see the sensitivity limits of this device and saw performance close to unity IDE but not full saturation. 

Along with the unchanged presence of an efficiency plateau at 7.4\textmu m, the main improvement in the SNAP and impedance-matched devices is the gain in SNR. To demonstrate the improvement in SNR, we reference the M60 detector used in Ref.~\cite{Colangelo2022} and measure it in the same cryostat with the same readout electronics as the other devices tested in this work. From Fig.~\ref{fig:PulseTraces}(a), adding the impedance matching taper in Device A improved the slew rate and voltage amplitude of the device, and raised the SNR. Adding the 2-SNAP in Device B further improves the amplitude and the SNR. Similarly, amplitude, slew rate and SNR are improved when comparing the 2-SNAP and 3-SNAP devices in Fig.~\ref{fig:PulseTraces}(b) to the reference device. To see if we could further improve SNR, we added a 120 MHz low pass filter and characterized the SNR in this configuration. The results are summarized in Table \ref{table:2}. All four devices tested in this work possess better SNR than the reference detector, with and without the low pass filter. It needs to be mentioned that the reference pulse presented in Fig.~\ref{fig:PulseTraces} was taken with additional room temperature amplification (ZFL-500+) as the pulse signal was too weak average without excess noise distorting the pulse shape. As such, the reference pulse is shown with the characterized gain of the amplifier removed from it. 

\begin{table}[ht]
\centering
\begin{tabular}{ |c|c|c|c| }
 \hline
  Device & No Filter (dB) & 120MHz Low Pass Filter (dB)\\
  \hline\hline
  Ref & 13 & 14\\
  A & 20 & 21 \\
  B & 24 & 25\\
  C & 27 & 26 \\
  D & 28 &  33 \\
  \hline
\end{tabular}
\caption{Summary of SNR improvement for all devices with and without a 120 MHz low pass filter.}
\label{table:2}
\end{table}
\begin{figure}[ht]
    \includegraphics[width=\linewidth]{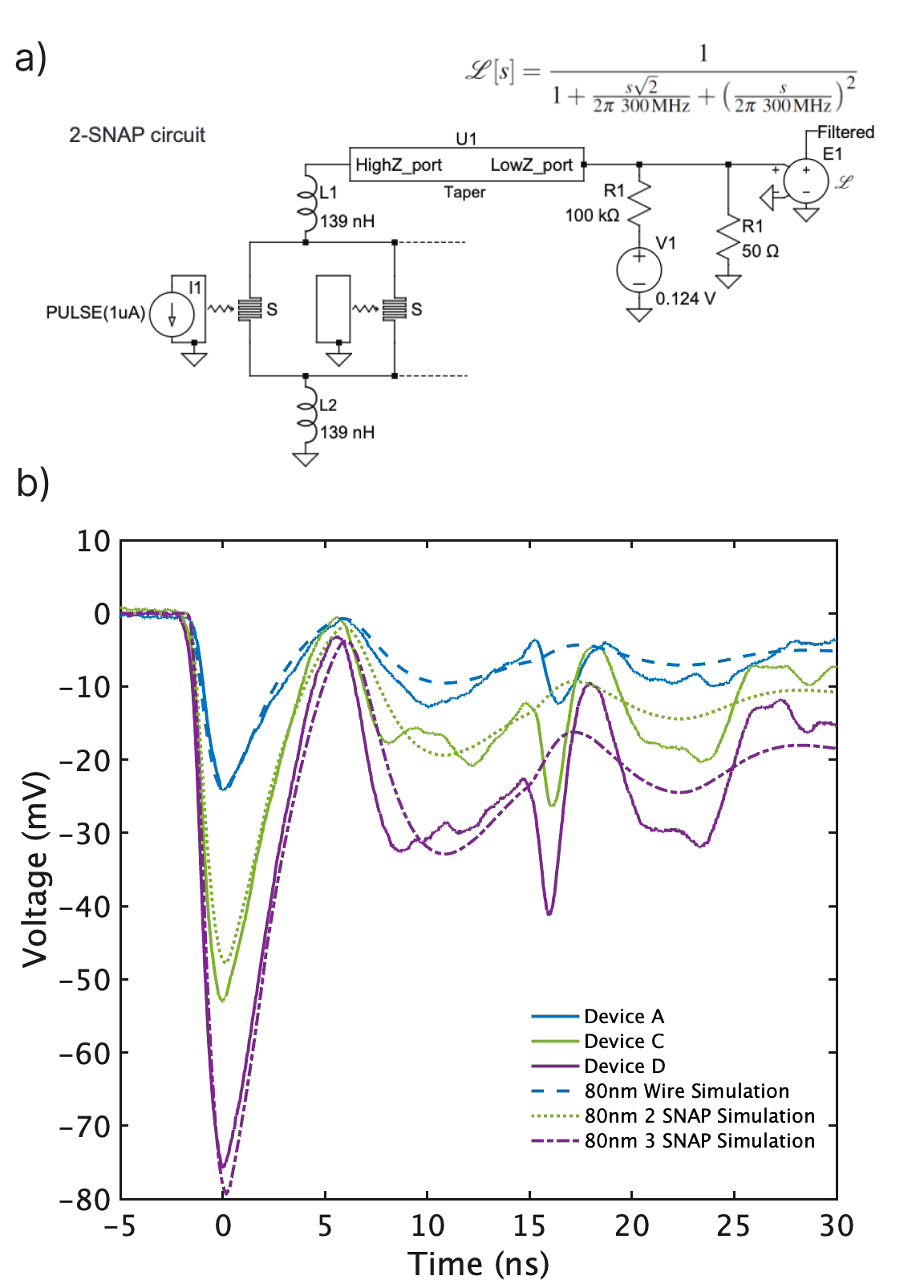}
    \caption{(a) Simulation circuit for the 2-SNAP device. (b) Simulated Pulse Traces for Devices A,C, and D at the bias currents specified in Fig.~\ref{fig:PulseTraces} for each respective device. We note that all devices line up closely with theoretical results.} 
    \label{fig:Simulations}
\end{figure}
To confirm our experimental result, we simulated the device response with SPICE. The photo-sensitive nanowires were modeled using Ref. \cite{berggren2018superconducting}, while the tapers were made of 300-section lossless transmission lines with variable inductance and capacitance following the Klopfenstein design with a $80\,\mathrm{MHz}$ bandwidth\cite{colangelo2023impedance}. Fig.~\ref{fig:Simulations}(a) shows the SPICE implementation of the 2-SNAP circuit simulation. The values for the inductances and bias were set to match the original layout. In Fig.~\ref{fig:Simulations}, we compare the simulated pulse response curves with the experimental data. Note that the simulation did not include any signal amplification or attenuation. For this reason, we scaled the simulated trace for the $80\,\mathrm{nm}$-wide wire to match the pulse amplitude of the corresponding measurement. We used the same scaling factor with the other curves. The simulated amplitude of the 2-SNAP and 3-SNAP are well in agreement with the experimental values, matching the measured SNR improvement. The slew rate of the rising edge was reproduced in the simulation by including a $300\,\mathrm{MHz}$ 2-pole low-pass filter. We attribute this filtering effect to the frequency-dependent attenuation of the readout circuit.   

\section{\label{sec:discussion}Discussion and Summary\\} 

In this work, we have demonstrated the use of impedance matching tapers and SNAP architecture to significantly improve the SNR performance of SNSPDs designed for the MIR. From our investigation we see that the addition of impedance matching tapers and SNAPs does not act as a detriment to the IDE at 7.4~\textmu m and likely maintains saturated IDE up until 10.6~\textmu m. Our experimental data was also largely in agreement with simulation demonstrating proper scaling with increasing number of parallel nanowires. This work ultimately illustrates the possibility of using such an architecture to optimize SNSPD development for the MIR and further. Additionally, this architecture offers the added benefit of increasing the active area of the detector while lowering the total kinetic inductance in the photosensitive area. In the future, we aim to further optimize the fabrication process with HSQ-defined SNSPDs by finding means of removing the HSQ soft mask from the WSi SNSPDs, as current techniques can damage the film and detector performance. While this did not degrade detector performance greatly at 7.4\textmu m it may hamper device performance at longer wavelengths. Future work will focus on the integration of this architecture into optical stacks and antennae coupling for optimized system detection efficiency along with  characterization with a calibrated light source. 

To summarize, we have taken two distinct device architectures, impedance matching tapers and SNAPs, and combined them to combat the low switching currents and correspondingly lower SNR in SNSPDs designed for the MIR. Our work confirms that the SNAP avalanche current mechanism holds up at lower bias currents and photon energies. Moreover, we show that impedance-matching tapers can replace the series inductors typically required in conventional SNAPs. This architecture acts as a promising platform for continuing to push SNSPD detector development for the abundance of applications in the MIR while simplifying the readout electronics.

\section{\label{sec:ack}Acknowledgments\\} 

S.P. was supported by a National Science Foundation Graduate Research Fellowship. M.C. was supported by the MIT Claude E. Shannon award. Part of this research was performed at the Jet Propulsion Laboratory, California Institute of Technology, under contract with the National Aeronautics and Space Administration (NASA - 80NM0018D0004). Support for this work was provided in part by the DARPA DSO Invisible Headlights program and the NASA ROSES-APRA program. The authors would also like to thank Sasha Sypkens for valuable feedback on the manuscript.

\bibliography{ref} 

\end{document}